# A novel fusion Python application of data mining techniques to evaluate airborne magnetic datasets

John Stephen Kayode and Yusri Yusup

*Abstract*— A novel fusion python application of data mining techniques (DMT) was designed and implemented to locate, identify, and delineate the subsurface structural pattern (SSP) of source rocks for the features of interest underlain the study area. The techniques of machine learning tools (MLT) helped to define magnetic anomaly source (MAS) rock and the various depths of these subsurface source rock features. The principal objective is to use straightforward DMT to locate magnetic anomaly features of interest that host mineralization. The required geo-referenced radiometric data, which facilitated the delineation of SSP, were sufficiently covered by combining the application of the Oasis Montaj© 2014 source parameter imaging functions. Relevance basic filtering techniques of data reduction were used to improve the signal-to-noise (S/N) ratio and hence automatically determine depths to the various engrossed features from gridded geo-referenced airborne magnetic datasets before the DMT application was performed. Geological source rock models (GSRM) (i.e., rock contacts, dykes) served as the delineated features based on their structural index (SI) values. The anomalies were perpendicularly oriented, with few inconsequential nonvertical features, and all were generally aligned in NNE-SSW and NE-SW directions. The DMT approach showed that magnetic anomaly patterns (MAP) control the SSP and the ground surface stratigraphy (GSS) on a geological time-scale (GTS) by fusing the subsurface gravitational structural features (SGSF) in the area. The DMT facilitated the determination of depths to these subsurface geological source rock features with a maximum depth of approximately 1.277 km using a 3x3 window size to map the concealed features of interest.

## I. INTRODUCTION

In recent years, extensive research has been reported on the applicability of state-of-the-art Machine learning techniques, (MLT) on varieties of environmental and geophysical prospection to enable advanced research and development of new methodologies. The use of present cutting edge python programming language (PPL) has significantly increased among the physical, and natural sciences, [1],[2],[3]. Extensive arrays of machine and deep learning libraries have been used for python-based statistical characterization [4] partly because its open-source toolbox is free and readily available to interested users. To support the quality contributions of previous reports on the applicability of MLT, this paper aims at bringing the approach of DMT for the location, identification, and delineation of SSP and the underlying source rocks in any geological environment. The foremost advantage is the lack of a requirement for prior knowledge on sophisticated computing with python. However, familiarity with computation and statistical analysis tools are an asset and beneficial in machine learning applications. It is anticipated that fusion python applications of the DMT would be fully developed for potential field theory methods of geophysical exploration given its user-friendly properties.

Many studies have been reported on the use of MATLAB and machine leaning techniques in diverse areas of geoscientific computing, e.g., [5], these techniques were initially applied to the statistical description of potential fields in the wavenumber domain to authenticate the reliability of the depths from a subsurface relief. Subsequently, [6] applied the Artificial Neural Network (ANN) to enhance the reliability, accurateness, and general quality of basement relief mapping. Following the success stories of these previous studies, [7] applied the Neural Network (NN) in an electromagnetic method to estimate the offset, depths, and conductivity area of a conductive object. Conversely, [8] used the Hopfield Neural Network (HNN) to process seismic data to overcome processing problems regularly encountered in seismic methods. In the same approach, [9] applied feed forward back-propagation NN techniques to analyze combined vertical electrical and induced polarization soundings to delineate coastal aquifers using synthetic datasets. In earlier reports of initial studies on the application of adaptable python programming for large datasets to map subsurface mineral potential, the present approach was not utilized. The significance of this technique involves the correct classification of the subsurface geological features (SGF) underlying the study area in the course of generating insightful knowledge of subsurface anomalies. Thus, this technique aid in better and smarter decisions for solid mineral prospection. The choice of the techniques was informed by the remarkable development and ease of a variety of

John Stephen Kayode was at Environmental Technology, School of Industrial Technology, Universiti Sains Malaysia, 11800, Pulau-Pinang, Malaysia. He is now with the Department of Research and Innovations, Institute of Hydrocarbons Recovery, Universiti Teknologi PETRONAS, Persiaran UTP, 32610 Seri Iskandar, Perak Darul Ridzuan, Malaysia.

Yusri Yusup is with the Environmental Technology, School of Industrial Technology, Universiti Sains Malaysia, 11800, Pulau-Pinang, Malaysia.

applications and the implementation of large datasets to map potential minerals in the study area [2],[3],[10], thereby validating and predicting the results.

The results of data analysis with the DMT helped to unveil and determine the subsurface linear features and the concealed subsurface structural anomalies (CSSA) beneath the study area, and this process is favorable for the identification of mineralization. Important geoscientific information on the geologic and lithological evolution of the subsurface structural frame work (SSFM) from in the study area is provided through the use of fusion python applications of NNs to automate and validate the output data results. Novel fusion python DMT along with other sophisticated established MLT were implemented using MATLAB and judged to be useful in Deep Learning technologies capable of training numerous networks based on the weights, the input, and the transfer function of the data being processed, which is consistent with a derivative function [2],[3],[11]. This attribute makes its selection suitable for the realization of the study set objectives [4].

To efficiently utilize the valuable subsurface precious ore rock bodies and other precious industrial minerals beneath the study area, the findings from this work are valuable for miners and other individuals involved in the mining sector of the economy. The principal motives behind a study of this type reside in the versatility of the results obtained from the data analysis, processes and validation. This information is exceptionally important to the Nigerian Geological Survey Agency (NGSA) and the Ministry of Solid Minerals for economic planning and policy implementation. Typically, the depths and other vital parameters serve as convenient information for the mining and industrial mineral developers [2],[3].

The data helped to generate features, derive impactful insight, visualize the information generated, and support accurate assessments of the mineral endowment of the area. The concept of data mining predominantly involves the use of the derived information from the analysis and measurements of large volume geo-referenced radiometric data. Given the principles underlining the deep learning applications, which are leveraged purely on the basis and strengths of statistical modeling, coupled with the massive amounts of geoscientific data generated in recent years, the applicability of python in MLT has become imperative as it is very easy to use and efficiently versatile in statistical computational analyses [4],[12],[13],[14],[15],[16],[17],[18],[19],[20].

When considering a reasonable amount of time to efficient handling large volumes of geo-referenced radiometric data without wasting time on designing and writing loops and codes while still implementing predictive deep learning techniques, python DMT with accelerated statistical computational capability tools serves as a better choice. In our modern society, python has provided a better supportive digital innovatory resource for modeling irrespective of scientific expertise given that the required codes or loops are built into the program. This simple interactive feature is due to the straightforward python program together with its friendly supporting statistical computational analysis tools. The fusion python DMT along with other refined and established MLT applications provide better choices for hugged statistical data analysis that will support the mining industries, academia, and government agencies. These technologies will provide the required knowledge to make well-informed decisions that could be realized globally [21],[22],[23],[24],[25].

## II. METHOD OF DATA ANALYSIS

Supervised machine learning (SML) tools (i.e., Fig. 1) using large-volume geo-referenced radiometric datasets to train the NN Pattern Recognition tools (NNPR) were developed to adequately map the SSP features beneath the study area. The MATLAB algorithm used is capable of training any network provided that the net input data, its weights, and the transfer functions include derivative functions. The method uses back propagation to compute the derivatives of the algorithm performance based on the weights and the bias attribute. The fascinating idea of developing the multifaceted tasks of DMT involves the ability of python, which is compatible with MATLAB, to automatically explore the large volume of the geo-referenced airborne magnetic datasets to establish the reliability of its patterns in relation to known variables, such as depth, size and nature of the subsurface anomaly. Therefore, this program could be used to automate and validate the output results for the geological features of interest. Typically, the challenges of inequalities in big datasets, such as random under-sampling or over-sampling classes of datasets or a mixture of both, are addressed during the preliminary data filtering processes and the use of relevant basic filtering tools for data reduction [26],[27],[28].

When certain sets of operations are developed and immersed into a MATLAB program to perform accurate assigned tasks of this magnitude, it could be enormously stressful, challenging, and psychologically demanding to delineate the SSP features of interest by generating new datasets from the existing primary data. The approach is to make predictions of the SSP occurrences for the MAS features beneath the study area. The enormous task relies on the capability of a novel fusion python application in machine learning to automate a complex statistical computational analysis of interconnected multicriterion steps for decision making, which enables the users to formulate acceptable accurate predictions of the output patterns generated. The technique was used to investigate the large volume of the primary geo-referenced radiometric airborne magnetic datasets using the assigned known variables connecting three foremost processing phases involved the following: (a) primary data exploration, (b) subsurface pattern identification and validation, and (c) development of the new datasets generated to make predictions [2],[3].

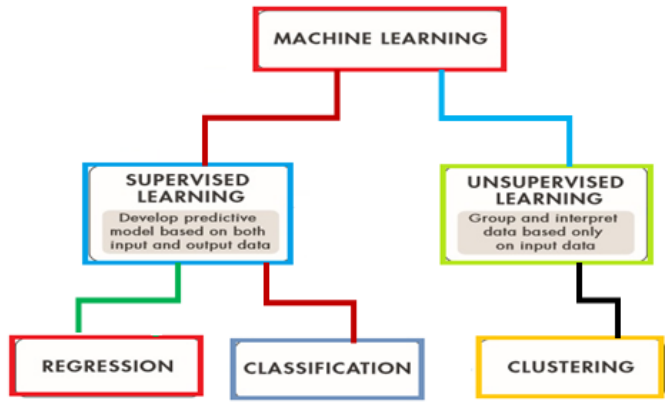

Fig. 1. Flowchart showing the processing steps for the machine learning technique. Supervised Machine learning tools, were selected and used for the large-volume geo-referenced radiometric datasets that trained the Neural Network Pattern Recognition tools developed to adequately map the subsurface structural pattern features beneath the study area.

### A. The primary data exploration

Exploration of primary data involved the initial data processing techniques and the use of relevant basic filtering data reduction techniques improved the signal-to-noise (S/N) ratio of the data. The process involves applications of the Euler deconvolution technique (EDT) to establish depths to the MAS using real geo-referenced radiometric datasets. However, some individual data points are possibly underrepresented compared to some neighborhood data points, yielding inequalities and inaccurate data points. This stage was performed to classify and sample the best significant variables, i.e., depth, and structural index, (SI) values, that defined the nature of the subsurface anomaly features with optimum correct classification using the principal component analysis (PCA), Table I. The phase involves the choice of simple shape geological models that represent the anomaly, thereby removing the effect of randomness and the undersampling and/or over-sampling of data classes [2],[29],[30],[31].

TABLE I
THE PRINCIPAL COMPONENT ANALYSIS (PCA) FOR THE GEO-REFERENCED RADIOMETRIC DATASETS.

|  | PC1 | PC2 | PC3 |
|---|---|---|---|
| X | 0.999972229326 | 0.999972229326 | -0.000366874925117 |
| Y | -0.00744343694204 | -0.00744343694204 | 0.000225500247564 |
| Z | 0.000368543267799 | 0.000368543267799 | 0.000225500247564 |
| Lambda | 255727523.207 | 253778529.367 | 1481.31440856 |

The principal component analysis (PCA) tool reported in Table I was used given its powerful applications in geostatistical data analysis. The tool offers desirable information from numerous intercorrelated variables for the features of interest that are quantitatively dependent. The aim of the PCA analysis is to extract useful information that will help to decrease the dimensionality of the datasets comprising a large number of independent variable features of interest while preserving the variations in the original complete geo-referenced radiometric datasets [32],[33],[34],[35].

*B. Pattern identification and validation*

In pattern identification and validation, it is essential to work with datasets lacking inequalities arising from bad data, aliasing, or underrepresented data points. To effectively deploy the fusion python in DMT, it is important to eliminate these problems before using the MATLAB algorithm to train the network of input data provided because almost all the standard MLT algorithms use equal data class distributions for python compatibility in MATLAB. Pattern identification and validation are involved in the selection of the most suitable modes that suit the features of interest to create the standard network that uses the input data from primary datasets and to train the data to compute the predictive data mining, thereby comparing their performance as the generated output [2],[3],[36],[37]. The large volume of primary data processed for data mining and pattern recognition satisfied the needs of the novel python fusion application, which automates complex statistical analysis of the interconnected complete datasets for pattern identification and validation. This feature enables acceptable correct predictions of the output patterns generated using the complete spatial randomness (CSR) tool presented in Table II. The tool helped to provide valuable insight into the features represented in the complete geo-referenced radiometric datasets.

TABLE II
COMPLETE SPATIAL RANDOMNESS,(CSR) FOR THE GEO-REFERENCED
RADIOMETRIC DATASETS.

| | |
|---|---|
| Lambda: | 9.85298857688e-005 |
| Clark and Evans: | 1.98524442595 |
| Skellam: | 1892451.75839 |

*C. Development of the model for new datasets*

To generate the desired prediction, the preceding stage, i.e., pattern identification and validation, is applied to the new data generated from the sampling of the primary datasets by the EDT to produce the expected features of interest at the output. The implementation involves computation of numerous statistical analyses to improve the quality of the geo-referenced radiometric data and the S/N ratio for novel fusion python applications. The data-mining concept has its most remarkable use in the management of vital information from the geo-referenced radiometric data with a long history of successes. Given that the principles are based on statistical characterization and modeling, it is possible to implement the techniques in any desired predictive machine learning model [2],[3],[38],[39].

Airborne geo-referenced radiometric data were collected by the Fugro Airborne Survey Services and Patterson Grant and Watson (PGW) using Proton Magnetometer Equipment with a fixed-wing aircraft, 3x Sintrex CS3 Cesium Vapor magnetometer, and FASDAS magnetic counter. The KING KR 405/KING KR 405B Radar altimeter and ENVIRO BARO/DIGIQUARTZ barometric altimeter served as the acquisition parameters. Data acquisition was contracted by the Nigerian Federal Government, and data were commercially available to interested users for the sole purpose of academic research through the Nigerian Geological Survey Agency(NGSA) (Fig. 2). Applications of various filtering processes for the removal of aliasing, deafening signals, data randomness, under-sampling of data and/or over-sampling of data classes, which leads to unnecessary/under classification, was performed to enhance the quality of the primary geo-referenced data with processes that utilized folding at a Nyquist frequency and a standard Kriging interpolation method of statistical computational analysis using the nearest neighbor tool [2],[3],[32],[34],[40],[41],[42],[43],[44]. The corrected geo-referenced data were projected via Universal Transverse Mercator (UTM) projection.

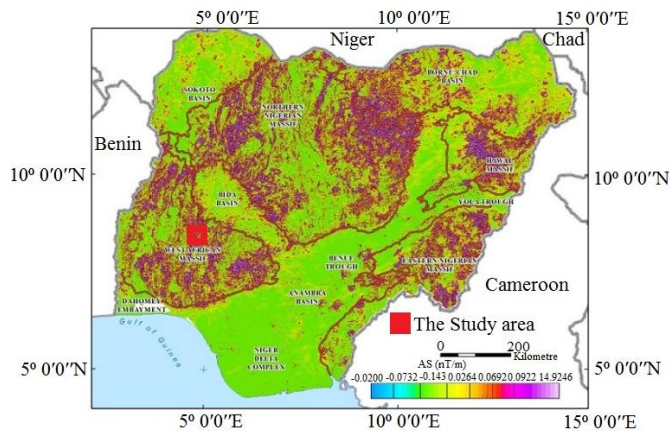

Fig. 2. Nigerian airborne magnetic map showing the study area modified by the Nigerian Geological Survey Agency.

## III. DISCUSSION OF RESULTS

The results reported in this paper reveal six structural indices (i.e., 0.5, 1.0, 1.5, 2.0, 2.5 & 3.0) along the 12 geomagnetic cross-sections with the application of different approaches from earlier studies with similar objectives, e.g., [2],[3],[40],[41],[42],[43],[44]. The best estimator for the range of values generated provides a minimum root-mean-square (RMS) error of zero and a maximum value of $8.9 \times 10^{-3}$. These values are therefore selected and grouped into the SI values, as shown in the results. The methods of magnetic anomaly interpretations adopted for the work helped to improve the map of the concealed subsurface structural anomaly features of interest in the Omu-Aran schist belt zone. Despite the fact that magnetic data are prone to noise, the choice of a small window size, i.e., the 3 x 3 window size that was implemented, together with the geo-referenced data enhancement filtering techniques applied provided high accuracy in the SI value determination for the source parameters. Excellent results were achieved through the application of the statistical computational analysis given that the geo-referenced airborne magnetic statistical properties greatly differ from the properties of deafening/redundant data points. Separation of the two is crucial as infinitesimal shifts in either of the x or y values could be detrimental to all datasets. The application of various filters to separate these aliasing or redundant data points becomes a straightforward task. This research could shed even more light on the anomaly patterns in the area that was researched [2],[3],[30],[38],[45],[46],[47],[48].

Accurate determination of depths to the various subsurface anomalies is the primary interest in all applications of magnetic methods of geophysical prospection. Depths to the subsurface geologic features (SGF) of interest are more valuable and superior to all other properties in any correct subsurface geologic structural interpretations (SGSI). The significance of depth determination cannot be over emphasized compared with the other properties of the magnetic anomaly. The varied nature of the subsurface geological terrains (SGT) in the study area as obtained from the EDT were delineated from the enhanced geo-referenced magnetic data with the primary intent of enhancing the signatures of the subsurface structural and lithological boundaries associated with joints, major and minor faults, rock contacts, and fractures beneath the Omu-Aran schist belt zone. The Euler solutions obtained were presented in the series of the East-West and North-South geomagnetic cross-sectional profiles plots shown in Figs. 3 to 14. The method exposed some interesting patterns in the distributions of the various structural subsurface geological anomaly features underneath the area. The details of these structural geologic features were presented as delineated along the 12 geomagnetic cross-sectional profiles [1],[2],[3]. The ranges of the SI values recorded in the region from the application of the Euler's solutions and implemented in the DMT are presented in Figs. 3 to 14. Nevertheless, detailed examinations of the clusters specified in the structural indices plotted, permitted the allotment of the SI value to a magnetic anomaly source model. The EDT is a complex task with an estimator of difficult depths, but it proven to be the best of all the depth determination techniques for gravity and magnetic prospection methods [40],[42],[43],[44]. The applied ESA achieved six various SI values for the subsurface geological features in the Omu-Aran Schist belt area with a maximum constraint depth of approximately 1.277 km. The different subsurface geological source (SGS) models of these features of interest, i.e., the rock contacts, dykes, sills, cylinders, pipes, and spheres, correspond to the SI values recorded, and all were defined by the DMT statistical computational analysis applied in the MATLAB algorithm [2],[3].

Given that real magnetic anomaly source bodies are more varied in nature than the simple poles and dipoles sources, the original magnetic field geo-referenced datasets are mostly noisy. Approximately 343,982 enhanced data points with up of 100-m interspacing between the data points were used with a 3 x 3 window size to sufficiently define the subsurface of the Omu-Aran region across a land area of approximately 30,568.7 km$^2$ [1],[2],[3]. The MLT computational statistical analysis approach was applied to facilitate the data mining process with the required primary geo-referenced data from the interpreted airborne magnetic datasets using simple computer codes explicable for the MATLAB algorithm to establish the NNP. The program used six input data (i.e., the SI values), trained the data using one hundred hidden neutrons, and delivered a six-output model with a higher degree of precision. The significance of the techniques involves the accurate classification of the subsurface geological structural features (SGSF) that host the minerals beneath the study area through the generation of insights into the subsurface anomalies, thereby aiding decision- and policy-makers on the solid mineral prospects in the area. The choice of the technique was informed by its easy implementation in statistical computation analysis and management of large volumes of geo-referenced datasets. Most importantly, this system can validate the results of the Euler solutions, thereby distinguishing this technology from most previous studies. Specifically, this process incorporated the MLT to characterize these subsurface anomaly features in the Omu-Aran schist belt zone.

### A. Application of Nearest Neighbor Statistical tool

The Nearest Neighbor Statistical (NNS) tool was performed for the pattern classification of airborne magnetic datasets by returning the mean distance between neighbors of each feature of interest in the geo-referenced data points that was sampled at 100-m intervals (Table III). The development of NNS in DMT allows complete analysis of geo-referenced datasets through the selection of a set of values per time that are independent of each other [30]. The machining learning algorithm employed a vector operation by taking a unit order representing the features on the geo-referenced radiometric datasets and applying the instructions to the numerous data points that are measured and are freely located within the study area at a given time. To use this tool, the mean distance (in this case 100 m) should not be less than the average for the theoretical random distribution. Otherwise, a clustering situation will occur. However, in this study, the subsurface features being analyzed are discrete or isolated within the study area. Thus, the location of all the features can be analyzed separately, i.e., independently of each other.

TABLE III
THE NEAREST NEIGHBOR STATISTICAL TOOL (NNS) WAS
IMPLEMENTED FOR PATTERN CLASSIFICATION OF THE
GEO-REFERENCED MAGNETIC DATASETS.

|  | \|Delta Z\| | Separation |
|---|---|---|
| 1%-tile: | 100 | 0.00822448730469 |
| 5%-tile: | 100 | 0.044075012207 |
| 10%-tile: | 100 | 0.0926742553711 |
| 25%-tile: | 100 | 0.276733398438 |
| 50%-tile: | 100 | 0.775947570801 |
| 75%-tile: | 100 | 1.89129638672 |
| 90%-tile: | 100 | 3.88102340698 |
| 95%-tile: | 100 | 5.78561401367 |
| 99%-tile: | 100 | 12.1034507751 |
| Minimum: | 100 | 0 |
| Maximum: | 100 | 147.794021606 |
| Mean: | 100 | 1.60750803826 |
| Median: | 100 | 0.775947570801 |
| Geometric Mean: | 99.9999999992 | N/A |
| Harmonic Mean: | 99.9999999993 | N/A |
| Root Mean Square: | 100 | 3.11433115672 |
| Trim Mean (10%): | 100 | 1.22533499715 |
| Interquartile Mean: | 100 | 0.870310791964 |
| Midrange: | 100 | 73.8970108032 |
| Winsorized Mean: | 100 | 1.2626223423 |
| TriMean: | 100 | 0.929981231689 |
| Variance: | 0 | 7.1149997361 |
| Standard Deviation: | N/A | 2.6673956842 |
| Interquartile Range: | 0 | 1.61456298828 |
| Range: | 0 | 147.794021606 |
| Mean Difference: | N/A | N/A |
| Median Abs. Deviation: | 0 | 0.6090259552 |
| Average Abs. Deviation: | 0 | 1.29439550074 |
| Quartile Dispersion: | 0 | N/A |
| Relative Mean Diff.: | N/A | N/A |
| Standard Error: | N/A | 0.00482446276908 |
| Coef. of Variation: | N/A | 1.65933582957 |
| Skewness: | N/A | 7.38378169144 |
| Kurtosis: | N/A | 141.94744609 |
| Sum: | 30568700 | 491394.309691 |
| Sum Absolute: | 30568700 | 491394.309691 |
| Sum Squares: | 3056870000 | 2964876.11211 |
| Mean Square: | 10000 | 9.69905855372 |

*B. Structural index (SI), results and validation*

*1) Plots of the SI results along the E-W geomagnetic cross-section profiles.*

Fig. 3 shows the plot of SI results with depths along the E-W geomagnetic cross-section of profile 1 with a maximum depth of slightly >1.0 km. The anomalies were densely populated between approximately 150 m, below the ground surface, and down to a depth of approximately 600 m. The anomalies population diminished as the depth increases from approximately 600 to 800 m but is sparsely placed from approximately 800 m to approximately 1.0 km as shown. However, inconsequential nonvertical structures and crosswise anomalies are recorded in some points as shown near the surface at approximately 50 m and close to the horizontal distance of approximately 120000 in UTM units. Here, 445 x 6 data points were considered for the python fusion statistical computational analysis with RMS errors of $0 \leq 8.8 \times 10^{-3}$ recorded for the profile.

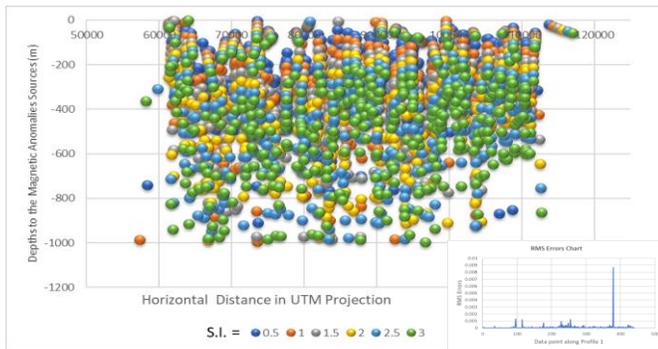

Fig. 3. Plot of SI results along E-W profile 1.

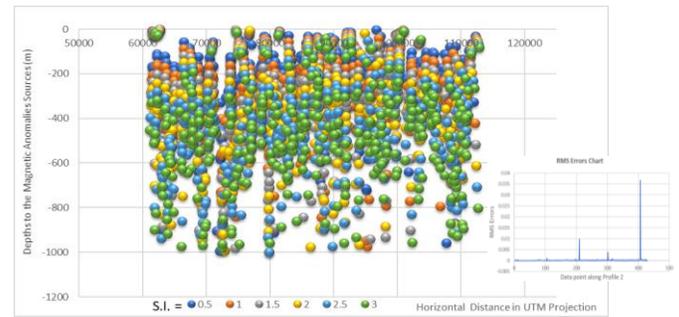

Fig. 4. Plot of SI results along E-W profile 2.

Fig. 4 shows some similarity in the distribution patterns of the recorded anomalies along profile 2. However, the maximum depth delineated along this geomagnetic cross-section profile to some extent is slightly deeper at approximately >1.1 km. The population anomalies were mostly concentrated between depths of approximately 100 m and 575 m with sparse distributions recorded between approximately 800 m and >1.0 km. The subsurface anomalies delineated along this profile are mostly vertical in orientation with the exception of very few nonvertical crosswise anomalies near the ground surface at a horizontal distance of approximately 112000 in UTM units. Here, 435 x 6 data points were used for the python fusion statistical computational analysis along this profile with recorded RMS errors of $0 \leq 3.7 \times 10^{-3}$.

The results recorded in the geomagnetic cross-section along profile 3 in Fig. 5 showed similar pattern distributions to those recorded in the preceding profile plots 1 and 2. The maximum depth of approximately 1.0 km was obtained with densely populated anomalies at depths between approximately 150 m and 500 m as shown. Following the same patterns as the previous 2 profiles, most of the recorded anomalies were vertically oriented with the exception of few insignificant nonvertical crosswise anomalies. The python fusion statistical analysis computerization utilized 450 x 6 data points with recorded RMS errors of $0 \leq 3.3 \times 10^{-3}$.

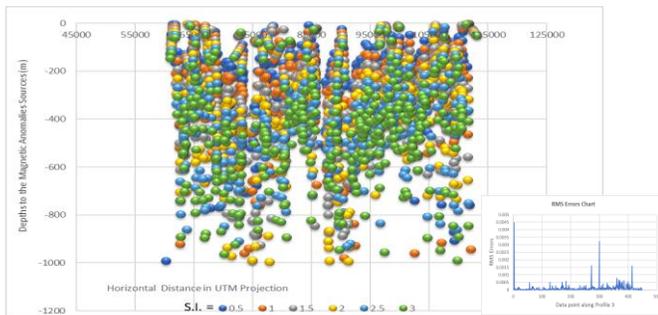

Fig. 5. Plot of SI results along E-W Profile 3

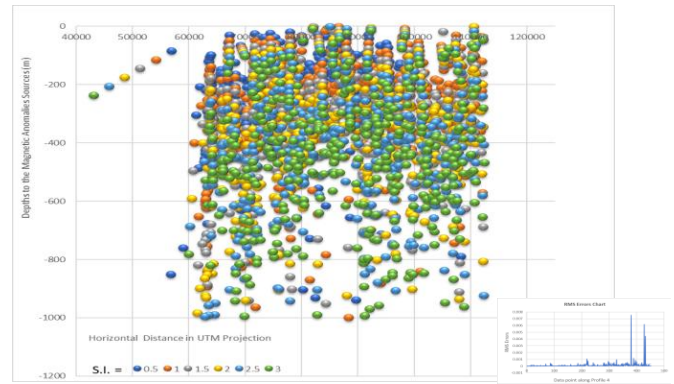

Fig. 6. Plot of SI results along E-W Profile 4.

The patterns of the anomalies recorded along the east-west geomagnetic cross-sectional profile 4 slightly differ from the preceding profiles 1 to 3. The nonvertical crosswise anomalies were delineated between depths of approximately 75 to 220 m along approximately 40000 to 58000 horizontal distance in UTM units were slightly reduced along approximate NNE-SSW directions. Thus, the profile is different as shown in Fig. 6. Nevertheless, the recorded anomalies are densely populated between depths of approximately 85 and 450 m. The profile produced further dense populated anomalies from depths of approximately 450 m to the maximum depth of slightly >1.0 km compared with that recorded in the previous geomagnetic cross-sectional profiles 1 to 3. The crosswise anomalies are reduced at an angle of approximately 45° and stretch from the deepest position at approximately 250 m to the shallow point at approximately 50 m below the ground surface. Here, 455 x 6 data points were implemented for the python fusion statistical computational analysis with RMS errors of $0 \leq 7.5 \times 10^{-3}$ recorded.

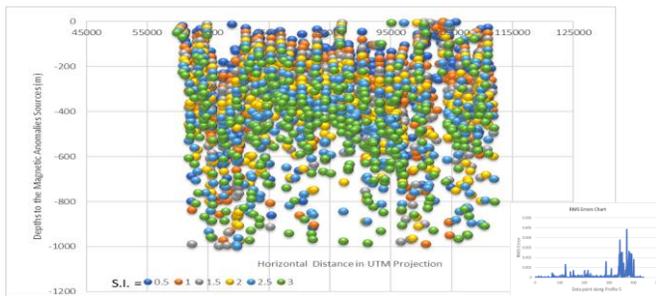

Fig. 7. Plot of SI results along E-W Profile 5.

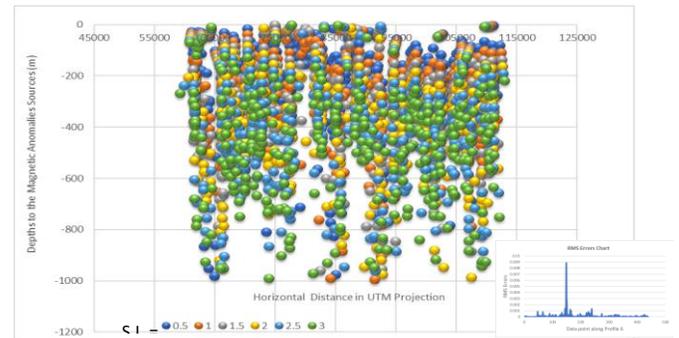

Fig. 8. Plot of SI results along E-W Profile 6.

The SI plots recorded along the E-W geomagnetic cross-sectional profile 5 as shown in Fig. 7 produced densely anomalies that are populated from depths between approximately 100 and 600 m with some few nonvertical crosswise anomalies. The patterns recorded along this profile followed the previously delineated results as enumerated in the past profiles with a maximum of slightly >1.0 km. Patterns were sparsely arranged around the central position between depths of approximately 700 m down to >1.0 km. Here, 443 x 6 data points were used for python fusion statistical analysis with the recorded RMS errors of $0 \leq 4.9 \times 10^{-3}$.

Fig. 8 presents the recorded patterns for the SI plots along the E-W Geomagnetic cross-section of profile 6 with numerous near surface nonvertical features. The population anomalies are also densely arranged from the ground surface down to a depth of approximately 700 m. Additional populations with nonvertical crosswise anomaly structures that dipped approximately in the NE-SW direction were delineated from depths of approximately 600 m to 1.0 km. The concentration of the anomalies lies between the ground surface down to depth of approximately 650 m and from the horizontal distance of approximately 60000 to <115000 in UTM units. The python fusion statistical computational analysis along the profile considered 445 x 6 data points with RMS errors of $0 \leq 8.9 \times 10^{-3}$ recorded.

### 2) Plot of the SI results along N-S geomagnetic cross-sectional profiles.

The general trends of the plot SI results for subsurface anomalies delineated along the N-S geomagnetic cross-sectional profiles showed complete divergence from that recorded along the E-W profiles. The population anomalies in the N-S geomagnetic cross-sectional profile 1 are concentrated along horizontal points between approximately 69000 and 70000 in UTM units. The shape resembles an elliptical shape with densely populated depths of approximately 0 to 600 m. Most of the features recorded are nonvertical in nature but are approximately along the NNE-SSW directions as shown in Fig. 9. The maximum depth of approximately >1.0 km was obtained with numerous crosswise anomalies that are sparsely scattered between depths of approximately 800 m and >1.0 km. A total number of 460 x 6 data points were sampled and used for the python fusion computerized statistical analysis with varied RMS errors of $0 \leq 1.2 \times 10^{-3}$.

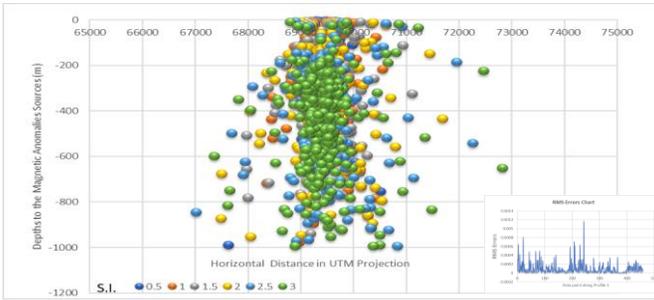

Fig. 9. Plot of SI results along N-S Profile 1.

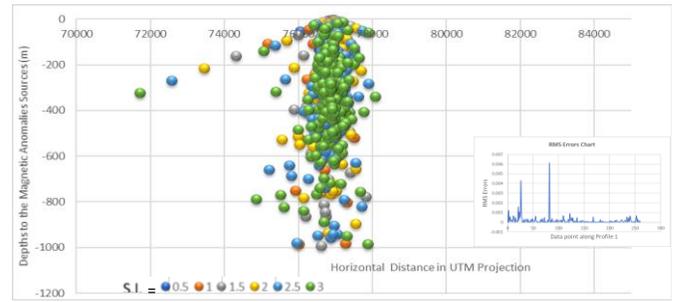

Fig. 10. Plot of SI results along N-S Profile 2.

The plot of SI results for the north-south geomagnetic cross-section along profile 2 is shown in Fig. 10. The results present a constricted shape that differs from the recorded results in N-S profile 1. The anomalies populated a narrow pipe-like shape with points that are concentrated along the horizontal position of approximately 77000 in UTM units at an approximate center of the profile. The anomalies spread from the ground surface downwards to a depth of approximately 800 m. The NNE-SSW structural orientation of the crosswise anomalies was recorded along the horizontal distance from approximately 72000 in UTM units at depth of approximately 300 m but become shallow as the horizontal distance increases towards the central point of the profile. The anomalies were compacted together along the midpoint of the profile with a maximum depth slightly greater than approximately 1.0 km. In total, 513 x 6 data points were used for the python fusion statistical computational analysis with recorded RMS errors of $0 \leq 6.1 \times 10^{-3}$.

Fig. 11 shows the plot of SI results for the anomaly patterns recorded along the N-S geomagnetic cross-sectional profile 3. The features of the anomaly population along the profile exhibited some resemblance to the shape recorded in the previous N-S profile 2 with further sparsely scattered crosswise anomalies located further downwards. The dense populations are concentrated at the midpoint of the profile, spreading from the ground surface to a depth of approximately 850 m. The crosswise subsurface anomaly structures are approximately located in the NNE-SSW direction as recorded in the first 2 N-S profiles. The python fusion statistical computational analysis used 415 x 6 data points with recorded RMS errors of $0 \leq 2.05 \times 10^{-3}$.

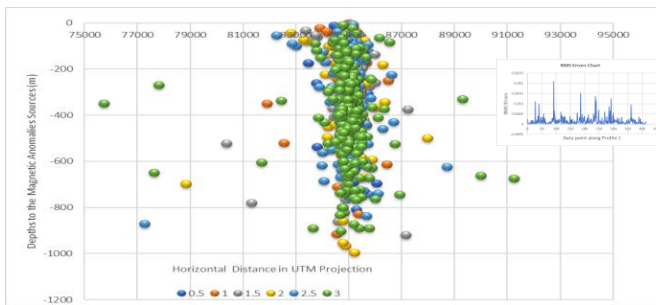

Fig. 11. Plot of SI results along N-S Profile 3.

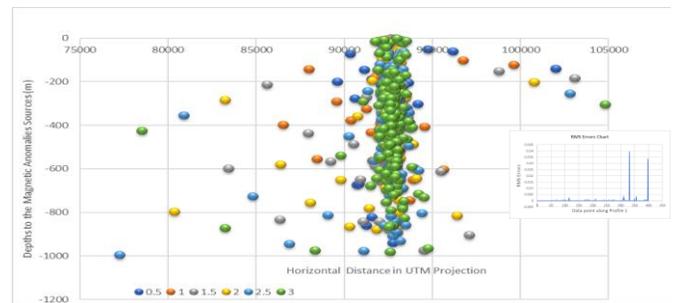

Fig. 12. Plot of SI results along N-S Profile 4.

The N-S geomagnetic cross-sectional profile 4, as shown in Fig. 12, is differs from that recorded in N-S profile 3, with the exception of more scattered diagonally arranged anomalies to the left-hand side of the centrally position and densely populated anomaly patterns along the horizontal position of approximately 90500 in UTM units. Only a minimal number of dispersed crosswise anomalies were recorded at the right-hand side of the figure in the opposite direction to those located at the left-hand side, scattering from slightly less than 91000 to 105000 in UTM units along the horizontal distance at an average depth of approximately 300 m. Nevertheless, the anomaly patterns were centrally located close together beneath the surface down to depths of approximately 950 m. Even so, the observed structures maintained the approximate NNE-SSW directions compared with other features previously recorded in the earlier profiles along the N-S geomagnetic cross-section. The python fusion statistical computational analysis used 403 x 6 data points with recorded RMS errors of $0 \leq 4.0$ x $10^{-3}$.

Results for the structural anomaly plots recorded along the geomagnetic cross-section of profile 5, as shown in Fig. 13, maintained the centrally populated features with a slightly broader shape than those achieved in the previous N-S profiles 2-4. The depth of the densely populated anomalies spread from the ground surface to approximately 550 m and sparsely populated from between approximately 600 m to the maximum depth of approximately 1.0 km. The subsurface structures could be observed between approximately 98000 to slightly greater than 102000 in UTM units along the horizontal distance in roughly NNE-SSW directions. Here, 438 x 6 data points were used for the python fusion statistical computational analysis with recorded RMS errors of $0 \leq 3.8$ x $10^{-3}$.

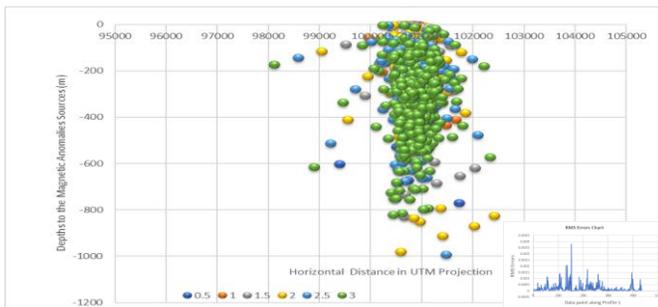

Fig. 13. Plot of SI results along N-S Profile 5.

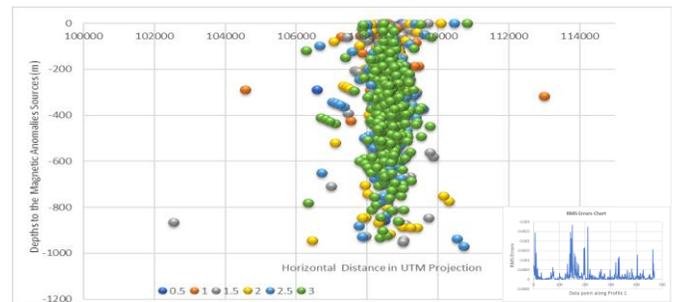

Fig. 14. Plot of SI results along N-S Profile 6.

The north-south geomagnetic cross-section along profile 6, as shown in Fig. 14, gives similar features obtained in the previous N-S profiles, except that it is shallower with maximum depth of approximately 980 m as against the >1.0 km maximum depths in all other previous profiles. On the other hand, the anomaly patterns recorded were centrally located between approximately 108000 and 110000 in UTM units along the horizontal distance with a few points that were uniquely placed in NW-SE directions. The subsurface structural features (SSF) were aligned between depths of approximately 200 m and 420 m and beneath the horizontal distance of 108000 in UTM units. The bulk of the anomalies are generally in a NNE-SSW orientation as recorded in the other earlier discussed N-S profiles. Here, 472 x 6 data points were used for the python fusion statistical computational analysis with the recorded RMS errors of $0 \leq 2.8$ x $10^{-3}$.

*3) Anomaly frequency of occurrence along the geomagnetic cross-section profiles against depths.*

MATLAB was used to compute the occurrence frequency and the width of these anomalies along the geomagnetic cross-section profiles. Fig. 15 shows the frequency plots along the geomagnetic cross-sections. Approximately 90% of the anomalies lie within approximately 0-650 m depth. On the other hand, the highest frequency of approximately 35 resides within the first 200-m depth. For depths greater than 650 m, the anomalies are sparsely placed. This finding may not be unconnected to the degree of weathering (i.e., both chemical and physical weathering) that occurred over the geological time-scale in the region that could have influenced the mineralization occurrences.

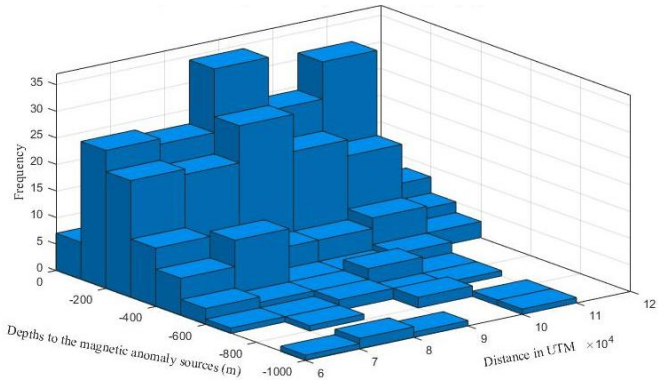

Fig. 15. Frequency plot against depths to the MAS along a typical geomagnetic cross-sectional profile.

## C. The generation of subsurface structural patterns (SSP).

The subsurface structural pattern generated from the fusion python statistical computational analysis is shown in Fig. 16. The inferred faults are indicated as F-F' by the assigned white lines shown in the figure. It is assumed that the subsurface magnetic anomaly features beneath the Omu-Aran region and the lithological characteristics in this part of the Nigerian Southwestern Precambrian Basement Complex control the topography. This notion was evidently observed by ground surface topography through the series of NNE-SSW and NE-SW orientations. These magnetic anomaly structural features will further control the patterns of the geologic anomalies and the subsurface strata in the era when the gravitational structures in the region were combined.

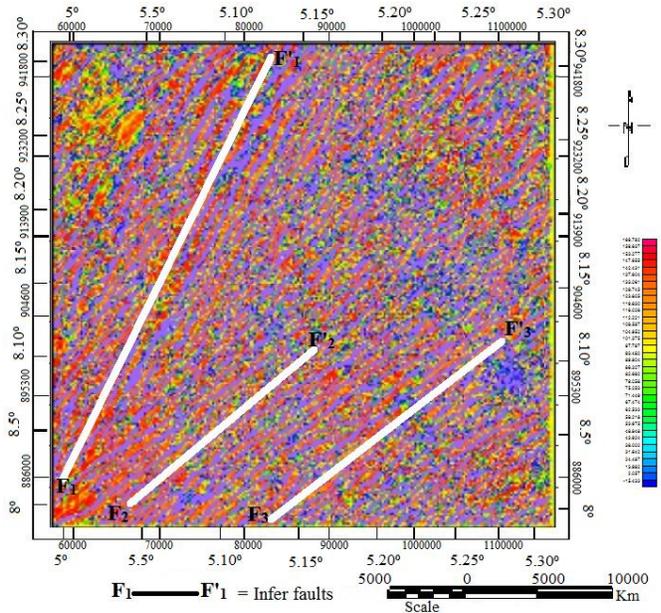

Fig. 16. The map of the study area showing the generated subsurface structural patterns (SSP) modified after [3].

## IV. CONCLUSION

The DMT approach demonstrated that the magnetic anomaly patterns (MAP) in the study area have a greater influence on the SSP, the ground surface stratigraphy (GSS) at the geological time-scale (GTS), and the merger of gravitational structural features (SGSF) in the area. Considering the applications of python statistical data computational analysis, it could be concluded that geological structural features (GSF) with the same trending patterns (STP) likely originated from the identical tectonic plate progression at some phase in the deformation stage, the GTS, and metamorphic maturity. The fusion python statistical computational analysis was applied to mine the airborne magnetic data and facilitated the delineation of depths to these subsurface geological source rock features with a maximum depth of approximately 1.277 km using a 3 x 3 window size.

### ACKNOWLEDGMENTS

The authors thank Universiti Sains Malaysia, Penang, Malaysia, for the doctoral research fellowship awarded in support of this work.

### COMPETING OF INTERESTS

The authors declare that they have no known competing financial interests or personal relationships that have or could be perceived to have influenced the work reported in this article.